# Local energy: a basis for local electronegativity and local hardness

Tamás Gál*

Quantum Theory Project, Department of Physics, University of Florida,

Gainesville, Florida 32611, USA

**Abstract:** The traditional approach to establishing a local measure of chemical hardness, by defining a local hardness concept through the derivative of the chemical potential with respect to the electron density, has been found to have limited chemical applicability, and has proved to be an unfeasible approach in principle. Here, we propose a new approach via a unique local energy concept. This local energy is shown to emerge from the Hamilton-Jacobi kind of construction of Schrödinger's quantum mechanics. It then leads to the concepts of a local chemical potential, i.e. negative of local electronegativity, and a local hardness just as the chemical potential and hardness are obtained from the energy, namely via differentiations with respect to the number of electrons. The emerging local hardness adds corrections to a recently proposed local hardness expression that has been found to be a good local measure of hardness for a series of atomic and molecular systems. These corrections become relevant for molecules with a large number of electrons. It is pointed out further that the definition of local softness that yields it as the Fukui function times the softness is not well-established, explaining recent observations of failure of this local softness concept as a proper local reactivity index for hard systems.

*Email: gal@qtp.ufl.edu

## I. Introduction

Chemical reactivity indicators have great significance in the study of chemical phenomena. The two most widely used such indices are the electronegativity [1] and the chemical hardness [2]. Although both indices have a history of many decades, even century, long, it was only after the birth of density functional theory (DFT) that they have got quantitative formulations, due to the works of Parr et al. [3,4], that are nowadays considered to be standards. The electronegativity has been identified as minus the electronic chemical potential [3], i.e. the derivative of the electronic ground-state energy with respect to the electron number $N$ in a fixed external potential setting, while the hardness has been given a definition as the second derivative of the ground-state energy with respect to $N$ [4]. DFT [5] does not play a direct role in these definitions, but it provides a natural background for their interpretation and evaluation, and especially for the introduction of a wide range of related chemical reactivity descriptors [6]. These quantitative definitions made it possible to place basic principles such as electronegativity equalization [7], hard/soft acid/base (HSAB) [2], and maximum hardness [8] on rigorous grounds [9]. Having reactivity indices that globally characterize electronic systems help chemists to predict the reactions between given species. Establishing local counterparts of these global quantities would enable chemists to make predictions regarding even the molecular sites a reaction eventually happens at. The idea of local reactivity indicators arose already in the works of Pearson, who proposed a local version of the hard/soft acid/base principle, too, and the aim of defining such indices has been of high importance of DFT [6]. The introduction of the Fukui function [10] as an indicator of the most reactive sites of molecules was a promising start, which was immediately followed by the proposal of a local softness concept [11], and of a local hardness concept [12]; so it seemed that Pearson's idea of a local HSAB might become a reality. However, the concept of local hardness was soon found to be undermined by ambiguity issues regarding its precise form [13,14], while very recently, even the local softness, which had not seemed to be a problematic quantity, has turned out to be a proper local measure of softness only for globally soft systems [15]. The problem with the local hardness has been found to be even more severe than "just" the question of finding a proper fixation of its ambiguity, as the traditional approach has proved to be inherently incapable of yielding a local hardness measure [16]. It is, therefore, necessary to establish a different theoretical framework for proper local measures of softness and hardness. Here, we will propose such a framework by showing that there exists a unique local energy that describes the energy density distribution of the $N$- electron system. A local hardness concept will then arise in the form of the second derivative of this local energy with respect to the electron number, while a local counterpart of the chemical potential, consequently of the electronegativity, will emerge as the first derivative with respect to the electron number.

## II. Background

Before introducing the new local concepts, it is worth highlighting the problems going along with the traditional definitions of local softness and hardness. The chemical potential, the negative of which has been identified as the electronegativity [3], is defined by

$$\mu = \left(\frac{\partial E[N,v]}{\partial N}\right)_v , \qquad (1)$$

with $E[N,v]$ denoting the ground-state energy of the $N$-electron system with external potential $v(\vec{r})$. $\mu$ appears as the Lagrange-multiplier corresponding to the fixation of electron number in the DFT Euler-Lagrange equation

$$\frac{\delta E_v[n]}{\delta n(\vec{r})} = \mu , \qquad (2a)$$

i.e.

$$\frac{\delta F[n]}{\delta n(\vec{r})} + v(\vec{r}) = \mu , \qquad (2b)$$

for the determination of the ground-state electron density in a given $v(\vec{r})$. $F[n]$ is a density functional giving the sum of the kinetic energy and the energy of interaction with each other of the electrons for a given density $n(\vec{r})$, and with it, the ground-state energy density functional $E_v[n]$ is given as $E_v[n] = F[n] + \int n(\vec{r}) v(\vec{r}) d\vec{r}$, whose minimization yields Eq.(2). As the derivative of $E_v[n]$ with respect to the density becomes constant only for the ground-state density, or linear combinations of ground-state densities, that corresponds to the given $v(\vec{r})$ (if the Lieb definition [17] of $F[n]$ is used), it is natural to consider $\frac{\delta E_v[n]}{\delta n(\vec{r})}$ as an $\vec{r}$-dependent chemical potential, which equalizes over space when the considered system of electrons reaches its ground state [3]. This then gives a theoretical formulation of the electronegativity equalization principle [7].

A further differentiation of the electronic ground-state energy with respect to the electron number yields the chemical hardness [4]

$$\eta = \left(\frac{\partial \mu[N,v]}{\partial N}\right)_v , \qquad (3)$$

while the inverse quantity defines softness,

$$S = \left(\frac{\partial N}{\partial \mu}\right)_v . \qquad (4)$$

By taking the derivative of Eq.(2) with respect to the electron number,

$$\int \frac{\delta^2 F[n]}{\delta n(\vec{r})\delta n(\vec{r}')} \left(\frac{\partial n(\vec{r}')}{\partial N}\right)_v d\vec{r}' = \eta , \qquad (5)$$

a hardness equalization principle can be established, paralleling the chemical potential equalization principle, since the left side of Eq.(5) is an $\vec{r}$-dependent quantity for a general $n(\vec{r})$, which becomes a constant only for ground-state densities [18]. This quantity has been introduced by Ghosh [13] as a potential local hardness measure,

$$\eta(\vec{r}) = \int \frac{\delta^2 F[n]}{\delta n(\vec{r})\delta n(\vec{r}')} \left(\frac{\partial n(\vec{r}')}{\partial N}\right)_v d\vec{r}' , \qquad (6)$$

and has been discovered to give a constant for ground-state densities, i.e. Eq.(5), by Harbola et al. [14]. The latter implies that Eq.(6) cannot be considered as a local measure of hardness, since the essence of a local hardness concept would be to locally characterize the hardness of molecular sites, differentiating between them point by point.

Eq.(6) was proposed as an alternative to the original local hardness formula of Berkowitz et al. [12],

$$\eta(\vec{r}) = \int \frac{\delta^2 F[n]}{\delta n(\vec{r})\delta n(\vec{r}')} \frac{n(\vec{r}')}{N} d\vec{r}' , \qquad (7)$$

which has been deduced from the more elementary definition [12]

$$\eta(\vec{r}) = \left(\frac{\delta \mu}{\delta n(\vec{r})}\right)_v . \qquad (8)$$

Eq.(8) is an appealing way to define the local hardness, since it appears as a natural local version of the hardness concept of Eq.(3), the electron density $n(\vec{r})$ being nothing else than the "local electron number". It has been recognized, however, that the fixed external potential constraint in Eq.(8) can be applied in several different ways, yielding several various definitions of local hardness [14,19]. Eq.(8) may lead to the constant local hardness of Eq.(6), too, by simply fixing $v(\vec{r})$ as one of the variables of $\mu[N,v]$, which actually is the most obvious choice. All choices necessarily have the common feature of yielding the (global) hardness via

$$\int \eta(\vec{r}) f(\vec{r}) d\vec{r} = \eta , \tag{9}$$

which can be seen by an application of the chain rule of differentiation. In Eq.(9), the notation

$$f(\vec{r}) = \left( \frac{\partial n(\vec{r})[N,v]}{\partial N} \right)_v \tag{10}$$

has been introduced, $f(\vec{r})$ being an essential quantity itself, which is not other than the Fukui function [10]. As has been pointed out by Ayers and Parr [20], most generally, Eq.(8) can be considered as a restricted derivative $\eta(\vec{r}) = \left. \frac{\delta \mu[N[n], v[n]]}{\delta n(\vec{r})} \right|_v$, where the $n(\vec{r})$-domain on which the differentiation is carried out is restricted by the constraint $v[n(\vec{r}')] = v(\vec{r})$ (see Sec.II of [21] for a general discussion of restricted derivatives). This implies a wide ambiguity of this local hardness concept [14,16].

However, although it is true that in the presence of constraints on the functional domain, more than one function $\frac{\delta A[\rho]}{\delta \rho(x)}$ will be capable of delivering the first-order changes of the given functional $A[\rho]$, this does not imply that all these functions will be derivatives that give the infinitesimal increment of $A[\rho]$ in accordance with the constraints properly [16]. To obtain a derivative that *in itself* is in accordance with the given constraint, the concept of constrained differentiation [21,22] has to be utilized, which in the case of the fixed-$v(\vec{r})$ constraint yields $\frac{\partial \mu[N,v]}{\partial N}$ as the only proper choice of $\left. \frac{\delta \mu}{\delta n(\vec{r})} \right|_v$ [16], i.e. the constant local hardness of Ghosh, Eq.(6). Another choice of $\left. \frac{\delta \mu}{\delta n(\vec{r})} \right|_v$, namely the unconstrained local hardness $\frac{\delta \mu}{\delta n(\vec{r})}$ of Ayers and Parr [23], may also be supported mathematically, if one considers the integrand in Eq.(9) as the local hardness instead of $\eta(\vec{r})$ of Eq.(8) solely; however, its evaluation is undermined in principle by the fact that $\mu[n]$ is given by $\mu[n] = \frac{\delta F[n]}{\delta n(\infty)}$ [16]. For further theoretical studies on the problematic nature of the local hardness concept of Eq.(8), we refer to [24,25]. We note here that the fact that other choices of Eq.(8) than the constant Eq.(6) is not allowed mathematically does not imply that such choices cannot have use as local indicators; it only implies that they will not measure local hardness. Eq.(7), e.g., has been shown recently [26] to be a local indicator of *sensitivity* towards perturbations, going against the essence of the concept of local hardness; see also [16], where Eq.(7) has been pointed out to emerge as a term of the "unconstrained local chemical potential".

Recently, it has been found by Torrent-Sucarrat et al. [15] that even the local softness as defined by [11]

$$s(\vec{r}) = \left(\frac{\partial n(\vec{r})}{\partial \mu}\right)_v \tag{11}$$

fails to be a proper local counterpart of softness: For globally hard molecules, it has been observed to give large values at hard sites, instead of the soft ones. This finding is very surprising considering the fact that the local softness concept of Eq.(11) had seemed to be well-established and without problems. The reasons to believe that Eq.(11) is a proper quantification of local softness were the following. (i) Eq.(11) can be deduced from the softness definition Eq.(4) in a very reasonable way, by simply replacing $N$ with its local counterpart $n(\vec{r})$. (ii) $s(\vec{r})$ of Eq.(11) is proportional with the Fukui function,

$$s(\vec{r}) = \left(\frac{\partial n(\vec{r})}{\partial N}\right)_v \left(\frac{\partial N}{\partial \mu}\right)_v = f(\vec{r})S \ , \tag{12}$$

while $f(\vec{r})$ was well-known to have largest values at the most reactive sites of molecules, which can usually be considered also the softest sites. (iii) In the theory of metals, Eq.(11) can be explicitly evaluated, giving the local density of states at the Fermi level [11], which may be considered as an indicator of local softness. However, the latter point is specific for metals, with a large number of electrons; in addition, besides the density of states term in $\left(\frac{\partial n(\vec{r})}{\partial \mu}\right)_v$, there is an additional term [27] that cannot be neglected for small metallic systems. The second point stops to be supportive if the Fukui function turns out not to indicate the reactive sites properly for hard molecules, as has been found in [15]. To see that point (i), in fact, does not give a support for Eq.(11) as a local counterpart of softness either, consider the following. Just as the softness has been introduced as the inverse of hardness, we may introduce the inverse chemical potential by

$$\mu^{-1} = \left(\frac{\partial N}{\partial E}\right)_v , \tag{13}$$

provided the function $E(N)$ is one-to-one, which is true in the case of real electronic systems, where the energy is monotonously decreasing with $N$. A corresponding local quantity may readily be defined by

$$\mu^{-1}(\vec{r}) = \left(\frac{\partial n(\vec{r})}{\partial E}\right)_v , \tag{14}$$

analogously with Eq.(11). This $\mu^{-1}(\vec{r})$ will then be proportional to $f(\vec{r})$,

$$\mu^{-1}(\vec{r}) = \left(\frac{\partial n(\vec{r})}{\partial N}\right)_v \left(\frac{\partial N}{\partial E}\right)_v = f(\vec{r})\mu^{-1} , \tag{15}$$

just as $s(\vec{r})$! That is, $\mu^{-1}(\vec{r})$ and $s(\vec{r})$ are the measures of the same local information, or one of them (or both) should be defined in some other way. (Note that this argument can be applied for higher derivatives of the energy with respect to $N$, too.) The observation of [15] indicates that Eq.(12) cannot be true; consequently, Eq.(11) needs to be corrected. We mention here that the finding of [15] is in accordance with earlier findings [28-30], which indicated that in certain types of reactions, the Fukui function does not function as expectations would dictate.

To overcome some of the above drawbacks, it has been proposed very recently [31] that instead of following the traditional way, Eq.(8), of defining a local hardness, through the replacement of $N$ in the hardness formula by the electron density, a local chemical potential $\mu(\vec{r})$ should be defined first, the derivative of which with respect to $N$ then would deliver a local hardness,

$$\eta(\vec{r}) = \left(\frac{\partial \mu(\vec{r})[N,v]}{\partial N}\right)_v . \tag{16}$$

$\mu(\vec{r})$ was defined in [31] as

$$\mu(\vec{r}) = \frac{n(\vec{r})}{N}\mu , \tag{17}$$

by utilizing the fact that the chemical potential emerges as the additive constant term in the $N$-conserving derivative [22] of the energy density functional $E_v[n]$ with respect to $n(\vec{r})$, i.e.

$$\mu = \frac{1}{N}\int n(\vec{r})\frac{\delta E_v[n]}{\delta n(\vec{r})}d\vec{r} . \tag{18}$$

Eq.(16) then yields

$$\eta(\vec{r}) = -f(\vec{r})\left(-\frac{\mu}{N}\right) + \frac{n(\vec{r})}{N}\left(\eta - \frac{\mu}{N}\right) , \tag{19}$$

which shows the appealing feature of being in a kind of inverse relation with the Fukui function, as both bracketed factors are non-negative. In addition, Eq.(19) integrates to the hardness, just as Eq.(17) integrates to $\mu$. As the Fukui function is known to be a proper local indicator of softness for soft systems, this inverse relation with $f(\vec{r})$ can be considered as a correct feature of $\eta(\vec{r})$ of Eq.(19). To exhibit the nature of this inverse relationship, integrate Eq.(19) over a given region of space,

$$\Delta\eta = -\Delta f\left(-\frac{\mu}{N}\right) + \frac{\Delta N}{N}\left(\eta - \frac{\mu}{N}\right). \tag{20}$$

It can be seen by dividing the electron cloud of the given molecule (with global properties $N$, $\mu$, and $\eta$) into arbitrarily small pieces containing the same "number" of electrons, $\Delta N$, that over these pieces the regional hardness values will be the additive inverses (times a constant) of the corresponding regional Fukui values. The fact that for hard systems, $f(\vec{r})$ fails to be a proper local softness index raises the question as to whether Eq.(19) is a good local hardness measure for hard systems, too. To examine this question, it is worth rewriting Eq.(19) in the form

$$\frac{\eta(\vec{r})}{\eta} + f(\vec{r})\left(-\frac{\mu}{N\eta}\right) = \sigma(\vec{r})\left(1 - \frac{\mu}{N\eta}\right), \tag{21}$$

which exhibits a relation between local quantities integrating to one, thereby making their comparison more feasible. The shape function $\sigma(\vec{r}) = \frac{n(\vec{r})}{N}$ has proved to be a relevant quantity in DFT [21,32,33]; also, it emerges as the leading term in gradient expansions of $f(\vec{r})$ [34]. It can be seen that for hard systems, with large $\eta$, the Fukui function gives a small contribution in Eq.(21), and $\eta(\vec{r})$ becomes proportional with the density $n(\vec{r})$. Since for hard systems, the harder sites are those with a larger number of electrons, this property of Eq.(19) supports it as a potential general local hardness measure. Numerical tests have also confirmed this [31], in particular for the critical test molecule benzocyclobutadiene, for which traditional definitions of $\eta(\vec{r})$ have failed. We should note here that, of course, the Fukui term in Eq.(21) can be neglected for large $\eta$ only if $\mu$ does not grow large proportionally to $\eta$. There have been recent studies proposing that the behaviors of $\eta$ and $\mu$ mirror each other [30,35]. However, one should not expect a simple, general proportionality. Further, the finite-difference approximation $\eta = I - A$ (which one may consider as a support for the similarity of $\eta$ and $\mu$, as usually $I \gg A$) is not satisfactory if one wishes to obtain a proper chemical hardness descriptor [36]. But for large $N$, Eq.(19) may raise questions, since it yields a proportionality of $\eta(\vec{r})$ and $\sigma(\vec{r})$ in general, if $\mu(N)$ is not such that $\mu(N) \propto \eta(N)N$ at least.

Since $\mu(\vec{r})$ of Eq.(17) has been obtained from an integral expression, it necessarily goes together with an ambiguity. That is, it is a question as to which of the choices

$$\mu'(\vec{r}) = \frac{n(\vec{r})}{N}\mu + z(\vec{r}) \tag{22}$$

is the best to define a local chemical potential, where $z(\vec{r})$ is any function that integrates to zero. Maybe there is a choice for $z(\vec{r})$ that yields a better concept of $\eta(\vec{r})$ than Eq.(19), giving a correction to that expression. Not to mention that no physical support has been provided for Eq.(17) as choice for a local chemical potential concept – it has been introduced as an auxiliary quantity for defining a good $\eta(\vec{r})$. In the following, we will base the concept of a local chemical potential onto physical grounds, originating it from a local energy. As a consequence, a correction to Eq.(19) will indeed arise, which should be taken into account in more precise local hardness calculations.

### III. Local energy

Our starting point to find the proper local energy, i.e. the energy density distribution of electrons, is the Schrödinger equation itself,

$$\left(\hat{T} + \hat{V}_{ee} + \hat{V}_{ne}\right)\psi(\vec{r}_1 s_1,...,\vec{r}_N s_N) = E\psi(\vec{r}_1 s_1,...,\vec{r}_N s_N) \,, \tag{23}$$

where $\hat{T} = \sum_{i=1}^{N} -\tfrac{1}{2}\nabla_i^2$, $\hat{V}_{ee} = \sum_{i<j}\frac{1}{|\vec{r}_i - \vec{r}_j|}$, and $\hat{V}_{ne} = \sum_{i=1}^{N} v(\vec{r}_i)$. Integrating Eq.(23) multiplied by $\psi^*(\vec{r}_1 s_1,...,\vec{r}_N s_N)$ in all the coordinates but one and summing it in the spin variables, or alternatively, acting on Eq.(23) with the operator

$$\hat{n}(\vec{r}) = \sum_{i=1}^{N} \delta(\vec{r}_i - \vec{r}) \tag{24}$$

of the density then taking the expectation value divided by $N$, gives

$$t(\vec{r}) + v_{ee}(\vec{r}) + v_{ne}(\vec{r}) = \frac{E}{N}n(\vec{r}) \,, \tag{25}$$

with

$$t(\vec{r}) = (1/N)\langle\psi|\hat{n}(\vec{r})\hat{T}|\psi\rangle \,, \tag{26a}$$

$$v_{ee}(\vec{r}) = (1/N)\langle\psi|\hat{n}(\vec{r})\hat{V}_{ee}|\psi\rangle \,, \tag{26b}$$

$$v_{ne}(\vec{r}) = (1/N)\langle\psi|\hat{n}(\vec{r})\hat{V}_{ne}|\psi\rangle \,. \tag{26c}$$

Note that Eq.(25) is the starting point for Bader's Atoms in Molecule theory [37]. Integration of Eq.(25) over the whole space gives $T + V_{ee} + V_{ne} = E$, and Eqs.(26) may be considered as a kinetic energy density (though with negative values at certain regions of space), an electron-electron interaction energy density, and an energy density of the interaction with the external potential, respectively [38]. On the basis of Eq.(25), a proportionality of the energy corresponding to a given segment of the molecule to the number of electrons contained in that segment can be concluded [37]. However, Eq.(25) gives even more: The expression on its right-hand side can be considered as a total energy density (or local total energy),

$$e(\vec{r}) = \frac{E}{N} n(\vec{r}) \;. \tag{27}$$

In addition, Eq.(27) is not just one of the many possible choices for a total energy density, differing by terms integrating to zero, but it is obtained from Eq.(23) just as the electron density is constructed from the wave function – from which, then, the electron number can be got back by a final integration, in the same way as Eq.(25) gives the total energy. That is, Eq.(27) gives *the* total energy density of electrons. Eq.(27) distributes the energy according to the electron distribution, more precisely, according to $n(\vec{r})/N$.

The above conclusion can be placed onto more elementary grounds. It is well-known [39] that by utilizing the explicit complex form of the wave function, $\psi(\vec{r}_1 s_1,...,\vec{r}_N s_N,t) = A(\vec{r}_1 s_1,...,\vec{r}_N s_N,t) e^{i\varphi(\vec{r}_1 s_1,...,\vec{r}_N s_N,t)/\hbar}$, the time-dependent Schrödinger equation decouples into a modified Hamilton-Jacobi equation,

$$\frac{1}{2m_e}\sum_{i=1}^{N}(\nabla_{\vec{r}_i}\varphi)^2 + \sum_{i<j}\frac{1}{|\vec{r}_i - \vec{r}_j|} + \sum_{i=1}^{N}v(\vec{r}_i) - \frac{\hbar^2}{2m_e}\sum_{i=1}^{N}\frac{\nabla_{\vec{r}_i}^2 A}{A} + \frac{\partial\varphi}{\partial t} = 0 \;, \tag{28}$$

with the quantum correction term $\frac{\hbar^2}{2m_e}\sum_{i=1}^{N}\frac{\nabla_{\vec{r}_i}^2 A}{A}$, and a continuity equation,

$$\frac{1}{m_e}\sum_{i=1}^{N}\nabla_{\vec{r}_i}\left(A^2 \nabla_{\vec{r}_i}\varphi\right) + \frac{\partial A^2}{\partial t} = 0 \;. \tag{29}$$

In Hamilton-Jacobi mechanics, the central quantity is Hamilton's principal function $S(q_1,...,q_n,t)$, which satisfies the Hamilton-Jacobi equation

$$H\left(q_1,...,q_n;\frac{\partial S}{\partial q_1},...,\frac{\partial S}{\partial q_n};t\right) + \frac{\partial S}{\partial t} = 0 \;. \tag{30}$$

$S(q_1,...,q_n,t)$ also has $n$ independent parameters, the $n$ constants of integration (plus one more, but irrelevant, additive constant). When the Hamiltonian does not depend explicitly on time, $S(q_1,...,q_n,t)$ can be written in the form $S(q_1,...,q_n,t) = W(q_1,...,q_n) - Et$, Eq.(30) becoming

$$H\left(q_1,...,q_n;\frac{\partial W}{\partial q_1},...,\frac{\partial W}{\partial q_n}\right) - E = 0 \ . \qquad (31)$$

As $W(q_1,...,q_n)$ solves Eq.(31), it is such that it yields the same value $E$ for any values of the $q_i$'s, when it is inserted in Eq.(31). In the time-independent case in quantum mechanics, $\varphi(\vec{r}_1 s_1,...,\vec{r}_N s_N,t)$ can similarly be written as $\varphi(\vec{r}_1 s_1,...,\vec{r}_N s_N,t) = \varphi(\vec{r}_1 s_1,...,\vec{r}_N s_N) - Et$. $\varphi(\vec{r}_1 s_1,...,\vec{r}_N s_N)$ corresponds to Hamilton's characteristic function $W(q_1,...,q_n)$; it is a function that yields the same energy value $E$ for any $(\vec{r}_i, s_i)$'s, but now together with the additional quantity $A(\vec{r}_1 s_1,...,\vec{r}_N s_N)$. Returning to the complex form $\psi(\vec{r}_1 s_1,...,\vec{r}_N s_N)$, which takes together $\varphi(\vec{r}_1 s_1,...,\vec{r}_N s_N)$ and $A(\vec{r}_1 s_1,...,\vec{r}_N s_N)$, it is clear that the time-independent Schrödinger equation,

$$-\frac{\hbar^2}{2}\sum_{i=1}^{N}\frac{\nabla_{\vec{r}_i}^2 \psi(\vec{r}_1 s_1,...,\vec{r}_N s_N)}{\psi(\vec{r}_1 s_1,...,\vec{r}_N s_N)} + \sum_{i<j}\frac{1}{|\vec{r}_i - \vec{r}_j|} + \sum_{i=1}^{N} v(\vec{r}_i) = E \ , \qquad (32)$$

can be considered as an equation that gives the same energy value $E$ for any $(\vec{r}_i, s_i)$'s for $\psi(\vec{r}_1 s_1,...,\vec{r}_N s_N)$'s that solve it. $\psi(\vec{r}_1 s_1,...,\vec{r}_N s_N)$ is not other than a generalized Hamilton's characteristic function, to incorporate quantum effects. For $\psi(\vec{r}_1 s_1,...,\vec{r}_N s_N)$'s that are not eigenfunctions, the left side of Eq.(32) will not give a constant over the space of $(\vec{r}_1 s_1,...,\vec{r}_N s_N)$'s; that is, it associates different energy values with different $(\vec{r}_1 s_1,...,\vec{r}_N s_N)$'s. Thus, finding the eigenfunctions of the Schrödinger equation amounts for finding those $\psi(\vec{r}_1 s_1,...,\vec{r}_N s_N)$'s for which the energy is constant for any configuration $(\vec{r}_1 s_1,...,\vec{r}_N s_N)$. This ensures that the total energy of the electrons is conserved during their motion in $v(\vec{r})$. It can be seen that the reason for which the Hamilton-Jacobi scheme is the proper choice to describe quantum mechanics is that its central quantity is such that associates the same total energy value for any point of the configurational space irrespective of the concrete position of the electrons, which is a must for quantum mechanics, where there are no well-defined trajectories of the moving particles. Of course, Eq.(32) implies the fulfillment of the (time-independent) continuity equation, so the imaginary terms cancel in Eq.(32).

If we restrict ourselves to $\psi(\vec{r}_1 s_1,...,\vec{r}_N s_N)$'s that satisfy the continuity equation, which are the only physically allowed wave functions, we have the real-valued energy

$$E(\vec{r}_1 s_1,...,\vec{r}_N s_N) \equiv -\frac{\hbar^2}{2}\sum_{i=1}^{N}\frac{\nabla_{\vec{r}_i}^2 \psi(\vec{r}_1 s_1,...,\vec{r}_N s_N)}{\psi(\vec{r}_1 s_1,...,\vec{r}_N s_N)} + \sum_{i<j}\frac{1}{|\vec{r}_i - \vec{r}_j|} + \sum_{i=1}^{N} v(\vec{r}_i) \qquad (33)$$

associated with every point of the configurational space. Eq.(33) can readily be extended to the time-dependent case, where a $t$ dependence appears besides the $(\vec{r}_i, s_i)$ dependencies, along with a $t$ dependence of $v(\vec{r})$. This will give, alternatively,

$$E(\vec{r}_1 s_1,...,\vec{r}_N s_N;t) = i\hbar \frac{1}{\psi(\vec{r}_1 s_1,...,\vec{r}_N s_N;t)} \frac{\partial \psi(\vec{r}_1 s_1,...,\vec{r}_N s_N;t)}{\partial t}.$$ Again, the real-valuedness of Eq.(33) is ensured by the requirement of the fulfillment of the continuity equation by $\psi(\vec{r}_1 s_1,...,\vec{r}_N s_N;t)$. Non-constant energy, corresponding to a $\psi(\vec{r}_1 s_1,...,\vec{r}_N s_N)$ that is not an energy eigenstate, of course, implies that the total energy of the electrons is not conserved during their motion in the given $v(\vec{r})$; this state should be considered as a (probability) superposition of energy eigenstates. The introduction of the energy concept of Eq.(33) is of course a matter of interpretation of quantum mechanics; like the most well-known, Bohmian alternative interpretation [40] (where the quantum correction term in Eq.(28) is associated with a physical meaning), or the hydrodynamical interpretation of Madelung [41] (see also [42]), it assumes new physics in addition to the standard, minimalistic, Copenhagen interpretation. We emphasize, however, that this is not the case above when concluding that Eq.(32) associates all points of configurational space with the same energy value. It is a simple fact that for (energy) eigenstates, (i) the left side of Eq.(32) is constant, and (ii) this constant equals the total energy of the $N$ electrons moving in $v(\vec{r})$.

It is worth mentioning that the connection between Schrödinger's quantum mechanics and Hamilton-Jacobi mechanics can be made even more explicit, by rewriting the time-dependent Schrödinger equation in the form

$$\frac{1}{2}\sum_{i=1}^{N}\left(\nabla_{\vec{r}_i}\Xi\right)^2 - \frac{i\hbar}{2}\sum_{i=1}^{N}\nabla_{\vec{r}_i}^2\Xi + \sum_{i<j}\frac{1}{|\vec{r}_i - \vec{r}_j|} + \sum_{i=1}^{N}v(\vec{r}_i) + \frac{\partial \Xi}{\partial t} = 0, \qquad (34)$$

where $\Xi = \varphi - i\hbar \ln A$ has been introduced [43]. The formal transition "from" classical mechanics to quantum mechanics is then apparent: The classical, real Hamilton's principal function should be corrected by $-i\hbar \ln A$, making it a complex number, plus the Hamilton-Jacobi equation should be corrected by the term $-\frac{i\hbar}{2}\sum_{i=1}^{N}\nabla_{\vec{r}_i}^2\Xi$. The so-called semiclassical limit, i.e. when $\hbar^2 \to 0$ is taken in Eq.(28), corresponds to neglecting the terms second-order in $\hbar$ in Eq.(34), while the classical limit is when $\hbar \to 0$. An immediate advantage of this scheme is that there arises no problem with the classical limit, the quantum action $\Xi$ simply reducing to $\varphi$, in contrast with $\psi = A e^{\frac{i}{\hbar}\varphi}$, where this cannot formally be done. Eq.(34) places the WKB approximation [39] (where Eqs.(28) and (29) with $\hbar^2 = 0$ are solved on the complex plane) on a more rigorous formal ground, too, since it simply corresponds to the new limit option offered by Eq.(34), namely neglecting the Laplacian term by $\hbar \to 0$ – which then requires the solution of the Hamilton-Jacobi equation on the complex plain. With $\Xi$, the quantum momenta of the particles arise as $\vec{p}_i(\vec{r}_1 s_1,...,\vec{r}_N s_N) = \nabla_{\vec{r}_i} \Xi(\vec{r}_1 s_1,...,\vec{r}_N s_N)$, in analogy with the Hamilton-Jacobi scheme. To obtain the momentum eigenstates, where the momentum of a particle is constant over the whole configurational space, one then needs to solve $\nabla_{\vec{r}} \Xi(\vec{r}s) = \vec{p}$, instead of $\frac{\hbar}{i} \nabla_{\vec{r}} \psi(\vec{r}s) = \vec{p} \psi(\vec{r}s)$. We note here that the arrangement of $\varphi$ and the quantum-mechanical $A$ into a complex number is a matter of choice – but, of course, with $\psi(\vec{r}s)$, we have the linear superposition of states, and a linear differential equation for it to solve.

With the position dependent energy concept of Eq.(33), now, we can easily justify Eq.(27) as *the* total energy density of electrons. For a ground state, we have that the system of $N$ electrons has a constant total energy during the motion of the electrons, no matter what the positions $(\vec{r}_i, s_i)$ of the $N$ electrons are. Then, if at a given $(\vec{r}_1 s_1,...,\vec{r}_N s_N)$, there is a higher probability of finding the $N$ electrons, the energy (probability) density will be proportionally higher there; namely, $e(\vec{r}_1 s_1,...,\vec{r}_N s_N) = E \psi^*(\vec{r}_1 s_1,...,\vec{r}_N s_N) \psi(\vec{r}_1 s_1,...,\vec{r}_N s_N)$. From this, Eq.(27) finally emerges through integrating in all coordinates but one. We note here that the well-known fact

$$\left( \frac{\delta E[N,v]}{\delta v(\vec{r})} \right)_N = n(\vec{r}) \qquad (35)$$

from perturbation theory nicely harmonizes with the constancy of the energy over $(\vec{r}_1 s_1,...,\vec{r}_N s_N)$, since if the energy is the same at every configuration of the electrons, the sensitivity of the ground-state energy to a small perturbation of $v(\vec{r})$ at a given point of space should be characterized by the "number" of electrons at that point. Of course, this is only for simple, multiplicative potentials; if the external field acting on the electrons differentiates, e.g., spin, the energy derivative with respect to $v_\sigma(\vec{r})$ will be determined by the "number" of spin-$\sigma$ electrons at $\vec{r}$, only.

## IV. Local electronegativity and local hardness

Having established the local energy Eq.(27), its derivatives with respect to the electron number, with $v(\vec{r})$ held fixed, naturally lead to the concepts of a local chemical potential, a local hardness, and local hyper-hardnesses. A local chemical potential can be defined as

$$\mu(\vec{r}) = \left(\frac{\partial e(\vec{r})[N,v]}{\partial N}\right)_v . \tag{36}$$

This $\mu(\vec{r})$ characterizes the change of the energy corresponding to a given point in a molecule due to a change in the electron number of the molecule, with nuclei kept fixed. The negative of $\mu(\vec{r})$ can be identified as a local electronegativity. Eq.(36) yields

$$\mu(\vec{r}) = f(\vec{r})\frac{E}{N} + \frac{n(\vec{r})}{N}\left(\mu - \frac{E}{N}\right) . \tag{37}$$

As can be seen, this expression corrects Eq.(17) by a term $\left(f(\vec{r}) - \frac{n(\vec{r})}{N}\right)\frac{E}{N}$, which integrates to zero. The second term of Eq.(37), on the other hand, gives a positive correction to the Fukui function term, since the bracketed factor of the shape function is non-negative. The latter is due to the convexity of the ground-state energy with respect to the electron number [44,45] and the fact that $-I \leq \mu \leq -A$; that is, $-\mu \leq I \leq -\frac{E}{N}$ while both the chemical potential and the energy are negative.

Applying the idea that has been applied in the case of Eq.(20), we may write Eq.(37) in a regional form,

$$\Delta\mu = \Delta f \frac{E}{N} + \frac{\Delta N}{N}\left(\mu - \frac{E}{N}\right) . \tag{38}$$

This shows that if we divide the electron cloud of a given molecule into arbitrarily small volumes containing the same number $\Delta N$ of electrons, over these pieces the corresponding regional chemical potentials will be in an "inverse" linear relation with the regional values of the Fukui function (the volumes can be infinitesimally small), due to the negativity of the energy. This may suggest to identify the local softness as the local electronegativity (times $-S/\mu$), since Eq.(37), i.e. minus the local electronegativity, has the Fukui function as the leading term, corrected by a term that gives the largest (negative) correction to $f(\vec{r})$ in the case of hard systems just where needed, namely where the density has the largest values. (Remember that $f(\vec{r})$, for hard systems, has been found to be large where $n(\vec{r})$ large [15], while these are the hard sites, requiring small local softness values.) In addition, $-\Delta\mu$ of Eq.(38) is in accordance with the general rule that small reactive sites should have a small local softness [30], since for a molecular region with small $\Delta N$, the second term in Eq.(38) is negligible if $\Delta f$ is (relatively) large. However, if we apply the qualitative "law" [30] that increasing hardness implies an increasing electronegativity, we find that $-\mu(\vec{r})$ cannot be a local softness indicator, since with $E/N$ (approximately) fixed, Eq.(37) gives a larger correction of the Fukui term for softer systems (smaller $\eta$, i.e. smaller $\mu$), where however the Fukui function is a proper local softness measure. All this indicates that $\mu(\vec{r})$ of Eq.(37) is an essentially new local reactivity index, an independent companion of local softness, or local hardness. To throw more light on the relation between $\mu(\vec{r})$ and $f(\vec{r})$, it is worth rewriting Eq.(37) in the form

$$\left(\frac{\mu(\vec{r})}{\mu} - \frac{n(\vec{r})}{N}\right) = \frac{E}{N\mu}\left(f(\vec{r}) - \frac{n(\vec{r})}{N}\right). \tag{39}$$

This shows that $\mu(\vec{r})$ is obtained through increasing the difference between the Fukui function and the shape function by a factor $\frac{E}{N\mu}$ (>1). This feature emerges from the fact that the local energy is proportional to the density.

Turning now to the second derivative of the local energy Eq.(27) with respect to $N$, the following local hardness formula arises:

$$\eta(\vec{r}) = \frac{\partial f(\vec{r})}{\partial N}\frac{E}{N} + 2\left(\frac{n(\vec{r})}{N} - f(\vec{r})\right)\frac{E}{N^2} + 2\left(\frac{n(\vec{r})}{N} - f(\vec{r})\right)\frac{-\mu}{N} + \frac{n(\vec{r})}{N}\eta. \tag{40}$$

Eq.(40) corrects Eq.(19) by the derivative of a term $\left(f(\vec{r}) - \frac{n(\vec{r})}{N}\right)\frac{E}{N}$ with respect to $N$, which integrates to zero. Combining Eq.(40) with Eq.(37), we obtain

$$\eta(\vec{r}) = -\mu(\vec{r})\frac{2}{N}\left(1 - \frac{\mu}{E/N}\right) + \frac{n(\vec{r})}{N}\left(\frac{2\mu}{N} - \frac{2\mu^2}{E} + \eta\right) + \frac{\partial f(\vec{r})}{\partial N}\frac{E}{N}, \tag{41}$$

which gives by what terms the term proportional to the local electronegativity $-\mu(\vec{r})$ needs to be corrected. One of the correction terms is proportional to the density, while the other one is proportional to the derivative $\frac{\partial f(\vec{r})}{\partial N}$ of the Fukui function, which has attracted much attention recently as a significant indicator of reactivity [46]. Note that the multiplier $\left(1 - \frac{\mu}{E/N}\right)$ of $-\mu(\vec{r})$ in Eq.(41) always lies between 0 and 1.

Dividing Eq.(40) by $\eta$ to have comparable local quantities, and applying the correlation rule between $\eta$ and $\mu$ mentioned earlier, we obtain that for large $\eta$, the first two terms of Eq.(40) can be neglected, leaving the terms already present in the local hardness expression of Eq.(19), except a factor 2 of the chemical potential term. This justifies Eq.(19) as a good approximate local hardness formula for hard systems. The factor 2 does not have much influence here, since for hard systems, that term does not affect the term $\frac{n(\vec{r})}{N}\eta$ significantly [31]. As Eq.(19) is a good local hardness measure for soft systems, being in an inverse relation with $f(\vec{r})$, Eq.(19) can be considered as a good approximation to local hardness in general (for $N$'s not very large, at least), which has been supported by numerical calculations, too [31]. The other question regarding the validity of Eq.(19) was its large-$N$ behavior, as it may give a simple proportionality with the density for large molecules. Examining Eq.(40), we can see that even if the second and the third terms become negligible for large $N$, the first term will still give a correction to the last term, since the energy increases comparably with $N$ (at least, for neutral or not very highly charged systems), hence the factor $E/N$ won't become small. Thus, for large systems, the direct use of Eq.(40) is required.

Eq.(40) is a unique local hardness concept. It tells us how much a given point of the molecule contributes to the global hardness, Eq.(3). This is a substantially different approach to defining a local hardness measure than the traditional approach of Eq.(8), which proposes a local hardness that indicates how much the chemical potential changes if we perturb the electron density at a point of the molecule, representing a local version of "how much the chemical potential changes if we perturb the electron number", which is the definition of global hardness. This is a reasonable approach, too; however, it is not capable to provide a local hardness indicator – instead, it yields an $\vec{r}$-dependent generalization of global hardness for non-stationary systems (in analogy with the $\vec{r}$-dependent electronegativity concept of Parr et al. [3]), which reduces to a constant quantity in space, $\eta$, for ground states. In contrast, the present definition of local hardness fits into the philosophy of "property density functions" [47], which is to find proper corresponding densities for global electronic properties to have local measures of the global quantities. Note that the local softness definition Eq.(11) follows this path, too; however, it has the problem of differentiating the density with respect to a quantity on which the density does not depend directly, only through a one-to-one dependence between the chemical potential and the electron number (consequently, Eq.(12), times $\frac{1}{S}\frac{\partial N}{\partial A}$, may be the density of any other quantity $\frac{\partial N}{\partial A}$ with an *A* in one-to-one correspondence with *N*). It is not surprising, though, that it is not possible to obtain a local softness concept by Eq.(11). The softness is the multiplicative inverse of hardness, for which a corresponding density function (hardness density) does exist, implying that the hardness is an extensive quantity (in a restricted sense) – but the multiplicative inverse of an extensive quantity won't be an extensive quantity! Fortunately, we do not need a local softness concept if we have a local hardness.

It is worth mentioning that one may introduce a density kernel and a chemical potential kernel by differentiating Eq.(27) with respect to the external potential and the density, respectively. In the case of the former, we obtain

$$n(\vec{r}',\vec{r}) = \frac{\delta e(\vec{r})[N,v]}{\delta v(\vec{r}')} = \frac{1}{N}n(\vec{r}')n(\vec{r}) + \frac{E}{N}\frac{\delta^2 E[N,v]}{\delta v(\vec{r})\delta v(\vec{r}')} \ , \qquad (42)$$

which is symmetric in its variables (!), and gives

$$\int n(\vec{r}',\vec{r})\,d\vec{r} = n(\vec{r}') \ . \qquad (43)$$

The other derivative is given by

$$\mu(\vec{r}',\vec{r}) = \frac{\delta e(\vec{r})[n,v]}{\delta n(\vec{r}')} = \frac{n(\vec{r})}{N}\frac{\delta E_v[n]}{\delta n(\vec{r}')} + \frac{E}{N}\left(\delta(\vec{r}'-\vec{r}) - \frac{n(\vec{r})}{N}\right) \ , \qquad (44)$$

and yields, on one hand,

$$\int \mu(\vec{r}',\vec{r})\,d\vec{r} = \frac{\delta E_v[n]}{\delta n(\vec{r}')} \ , \qquad (45)$$

(i.e. the $\vec{r}$-dependent, generalized chemical potential of Parr et al.), and on the other hand,

$$\int \mu(\vec{r}',\vec{r})f(\vec{r}')\,d\vec{r}' = \mu(\vec{r}) \ . \qquad (46)$$

The chemical potential itself is obtained as

$$\mu = \iint \mu(\vec{r}',\vec{r})f(\vec{r}')\,d\vec{r}'d\vec{r} \ , \qquad (47)$$

while the density kernel integrates to the electron number,

$$N = \iint n(\vec{r}',\vec{r})\,d\vec{r}'d\vec{r} \ . \qquad (48)$$

We note that as in Eq.(44), it is the only way to take the derivative of $E[N[n],v[n]]$, hence of $e(\vec{r})$, with respect to the density [16], similar to the case of $\mu[N[n],v[n]]$. One may obtain corresponding kernels with $\mu(\vec{r})$, too. A Fukui kernel may be introduced by

$$f(\vec{r}',\vec{r}) = \frac{\delta\mu(\vec{r})[N,v]}{\delta v(\vec{r}')} = \frac{n(\vec{r}')}{N}f(\vec{r}) + \frac{n(\vec{r})}{N}f(\vec{r}') - \frac{n(\vec{r}')}{N}\frac{n(\vec{r})}{N} + \frac{E}{N}\frac{\delta^2\mu[N,v]}{\delta v(\vec{r})\delta v(\vec{r}')} + \frac{1}{N}\left(\mu - \frac{E}{N}\right)\frac{\delta^2 E[N,v]}{\delta v(\vec{r})\delta v(\vec{r}')} \;,$$

(49)

which is symmetric in its variables, and integrates to the Fukui function in its second variable. $f(\vec{r}',\vec{r})$, of course, can be obtained as the derivative of $n(\vec{r}',\vec{r})$ with respect to $N$, due to the interchangeability of the differentiations with respect to $v(\vec{r}')$ and $N$. A hardness kernel, on the other hand, may be defined as

$$\eta(\vec{r}',\vec{r}) = \frac{\delta\mu(\vec{r})[n,v]}{\delta n(\vec{r}')} \;, \tag{50}$$

which integrates to the generalized hardness concept of Eq.(6) in its second variable, while in its first variable, integrates to the local hardness when multiplied by the Fukui function.

The spin-polarized versions of the above local chemical potential and local hardness concepts can be obtained readily, in both representations of spin-polarized DFT. The spin-polarized local chemical potentials emerge via differentiation of the local energy with respect to the corresponding spin particle numbers. I.e.,

$$\mu_\sigma(\vec{r}) = \left(\frac{\partial e(\vec{r})[N_\uparrow,N_\downarrow,v,B]}{\partial N_\sigma}\right)_{N_{\bar\sigma},v,B} = f_\sigma(\vec{r})\frac{E}{N} + \frac{n(\vec{r})}{N}\left(\mu_\sigma - \frac{E}{N}\right), \quad \sigma=\uparrow,\downarrow \;, \tag{51}$$

$$\mu_N(\vec{r}) = \left(\frac{\partial e(\vec{r})[N,N_s,v,B]}{\partial N}\right)_{N_s,v,B} = f_N(\vec{r})\frac{E}{N} + \frac{n(\vec{r})}{N}\left(\mu_N - \frac{E}{N}\right) , \tag{52a}$$

$$\mu_S(\vec{r}) = \left(\frac{\partial e(\vec{r})[N,N_s,v,B]}{\partial N_s}\right)_{N,v,B} = f_S(\vec{r})\frac{E}{N} + \frac{n(\vec{r})}{N}\mu_S \;, \tag{52b}$$

with $f_\sigma(\vec{r})$, $f_N(\vec{r})$, $f_S(\vec{r})$, and $\mu_\sigma$, $\mu_N$, $\mu_S$ denoting the derivatives of the density and the energy with respect to $N_\sigma$, $N$, $N_s$. The latter are the spin-polarized chemical potentials, while the one-indexed spin-polarized Fukui functions are related to the usual, two-indexed spin-polarized Fukui functions [48] by

$$f_\sigma(\vec{r}) = f_{\uparrow,\sigma}(\vec{r}) + f_{\downarrow,\sigma}(\vec{r}) \;, \tag{53}$$

$$f_N(\vec{r}) = f_{NN}(\vec{r}) \;, \qquad f_S(\vec{r}) = f_{NS}(\vec{r}) \;. \tag{54}$$

As can be checked, Eq.(51) yields a correction of the corresponding expression in [31] by a term $+\left(f_\sigma(\vec{r}) - \frac{n(\vec{r})}{N}\right)\frac{E}{N} + \frac{n_{\bar\sigma}(\vec{r})}{N_{\bar\sigma}}\mu_\sigma$. The spin-polarized local hardnesses $\eta_{\sigma\sigma}(\vec{r})$, $\eta_{\sigma\bar\sigma}(\vec{r})$, and $\eta_{NN}(\vec{r})$, $\eta_{NS}(\vec{r})$, $\eta_{SN}(\vec{r})$, $\eta_{SS}(\vec{r})$ can be obtained by differentiation of $\mu_\sigma(\vec{r})$, and $\mu_N(\vec{r})$, $\mu_S(\vec{r})$ with respect to $N_\sigma$, $N_{\bar\sigma}$, and $(N, N_s)$, $(N, N_s)$, respectively.

## V. Conclusion

It has been pointed out that on the basis of the Schrödinger equation, a unique local energy concept of electron systems can be established. This is supported by the fact that non-relativistic quantum mechanics has a Hamilton-Jacobi-kind structure. This local energy shows that the total energy of an electron system is distributed according to the density distribution of its electrons. The derivatives of the local energy with respect to the electron number deliver a local chemical potential, a local hardness, and local hyperhardnesses, in a way the derivatives of the total energy yield the corresponding global reactivity indicators. It is natural to identify the negative of the emerging local chemical potential as the local electronegativity, just as the electronegativity is identified with minus the chemical potential. These local indices give how much a given point of the molecule contributes to the corresponding global quantities. The obtained local hardness provides a sound theoretical background for an earlier proposed local hardness expression that has been found to be a good local measure of hardness for a series of electronic systems. That expression can be considered as a good approximation of the new formula for systems with a not very large number of electrons. It has been pointed out further that the traditional concept of local softness is not well established, explaining its recently found failure for hard molecules. This can be attributed to the fact that the local softness cannot be defined as a density distribution of softness, since the latter is not an extensive quantity, being the multiplicative inverse of a property, the hardness, that emerges as an extensive quantity. Hence, it is the local hardness which should be used to describe both the hard and the soft sites of a molecule.

**Acknowledgments:** The author acknowledges grants from the Fund for Scientific Research – Flanders (FWO) and the U.S. Department of Energy TMS program (Grant No. DE-SC0002139).